\documentclass{proclund}

\begin{document}
\pagestyle{prochead}

\title{KINEMATICAL vs DYNAMICAL RELATIVISTIC EFFECTS IN 
$A(\vec{e},e'\vec{p})B$}
\author{M.C. Mart\'{\i}nez$^{1}$, J.A. Caballero} 
\affiliation
{{Departamento de F\'{\i}sica At\'omica, Molecular y Nuclear, 
    Universidad de Sevilla, Apdo. 1065, 41080 Sevilla, Spain\\}}
\author{T.W. Donnelly}
\affiliation
{{Center for Theoretical Physics, Laboratory for Nuclear Science and
    Department of Physics, Massachusetts Institute of Technology,
    Cambridge, MA 02139, USA}}


\begin{abstract}
The relativistic mean field approach is applied to the description of
coincidence $A(e,e'p)B$ reactions where both the incident electron beam and
knockout proton are polarized. Effects introduced by the dynamical
enhancement of the lower components of the bound nucleon wave function
are analyzed within RPWIA 
for the polarized responses and transferred polarizations. 
Results obtained by projecting out the negative-energy components
are also compared with various nonrelativistic reductions.
\end{abstract}
\maketitle
\setcounter{page}{1}

\section{Introduction}

As is well known, over the years quasielastic coincidence $(e,e'p)$ 
reactions have
provided one of the most powerful tools to study
nuclear structure. In particular, single-particle properties such as
momentum distributions and spectroscopic factors corresponding to
nucleon shells in the vicinity of the Fermi level have been
extracted from the analysis of these processes. Moreover, $(e,e'p)$
reactions have also clearly proved the limits of the single-particle 
approach for nuclei on which the mean field approximation is based.
When polarization degrees of freedom are involved,
a much richer variety of polarization observables becomes accessible.
These contain in general
interferences between the various amplitudes and consequently a
complete decomposition into the electromagnetic matrix elements can in
principle be achieved. In the case of final-state nucleon
polarization measurements, i.e., $A(\vec{e},e'\vec{p})B$ processes, the
differential cross section can be written as~\cite{bof93}
\begin{equation}
\frac{d\sigma}{d\varepsilon_ed\Omega_ed\Omega_N}=\sigma_0[1+\vec{P}\cdot\vec{\sigma}+h(A+\vec{P}'\cdot\vec{\sigma})],
\end{equation}
where $\sigma_0$ is the unpolarized cross section, $h$ is the incident
electron helicity, $A$ denotes the electron analyzing power, and
$\vec{P}$ ($\vec{P}'$) represents the induced (transferred)
polarization. Note that $\vec{P}$
only depends on the outgoing nucleon
polarization, whereas $\vec{P}'$ becomes accessible when
the outgoing proton and electron beam polarizations are both
measured. The conventional three perpendicular
directions chosen to specify the recoil
nucleon polarization are given by $\vec{l}$ (parallel to the momentum 
$\vec{p}_N$ of the outgoing nucleon), $\vec{n}$ (perpendicular to the plane 
containing $\vec{p_N}$ and the transfer momentum $\vec{q}$), and $\vec{s}$ 
(determined by $\vec{n}\times \vec{l}$). In coplanar kinematics, 
the only surviving components are $P_n$, $P'_l$ and $P'_s$. Moreover,
the induced polarization $P_n$ is zero when final state
interactions (FSI) between the outgoing nucleon and the residual
nuclear system are neglected~\cite{pick8789}.

The transfer polarization components $P'_l$ and $P'_s$ may provide valuable
information on the nucleon form factors~\cite{malov}. In the case of 
electron-nucleon scattering one gets a close
relationship between the nucleon form factors and the polarization 
transfer components~\cite{arn81}
\begin{equation}
\frac{P'_s}{P'_l}=-\frac{G_E}{G_M}
[\tau(1+(1+\tau)\tan^2\frac{\theta_e}{2})]^{\frac{1}{2}},
\label{eq1}
\end{equation} 
with $\tau=\frac{|Q^2|}{4M_N^2}$, being $Q^\mu$ the transfer
four-momentum, and
$\theta_e$ the electron scattering angle. It is important to remark
that the above relationship is only strictly correct for electron
scattering from a free nucleon. In the case of bound nucleons
the polarization ratio should be evaluated within the scheme of a
particular nuclear model, and thus eq.~(\ref{eq1}) only holds approximately.
In spite of this, the `polarization' technique to determine the
nucleon form factors, presents clear
advantages compared with the usual Rosenbluth separation method.
It does not require one to vary the beam energy and/or the spectrometer
angle, thus eliminating the systematic uncertainties that make it so difficult 
to extract $G_E$ at high $Q^2$ within the Rosenbluth method.

Furthermore, recoil polarization calculations for low/medium missing momenta 
have also proven to be relatively insensitive
to different ingredients in the description of
the reaction mechanism, namely off-shell
ambiguities and optical potentials used to describe FSI~\cite{kel97}. 
These results allow one to consider the
polarization technique to be a promising probe in studying 
the behaviour of the nucleon form factors.

In recent years there has been a concerted experimental
effort to shed some light on the issue of the possible modification of the
form factors of the nucleons inside the nuclear medium.
High precision polarization transfer measurements on
complex nuclei have recently been presented by Malov 
et al. in $^{16}O(\vec{e},e'\vec{p})^{15}N$~\cite{mal00},
and by Dieterich et al. 
in $^4He(\vec{e},e'\vec{p})^3H$~\cite{die01}.
Although the general conclusions in both experiments are not free from
ambiguities due to experimental uncertainties, the authors in~\cite{die01}
show that standard nonrelativistic 
calculations are in clear disagreement with the 
experimental data. This result constitutes a strong indication of
the necessity for
a fully relativistic calculation in order to describe the spin transfer 
observables.

The relativistic mean field approach has been 
used recently to evaluate several electron scattering observables. These
have been successfully compared with experimental data for
transferred and 
induced polarizations~\cite{udi00}, as well as for
unpolarized observables: the interference
transverse-longitudinal response $R^{TL}$, left-right 
asymmetry $A^{TL}$~\cite{udi99}, single-particle momentum 
distributions~\cite{udi96} and spectroscopic 
factors~\cite{udi93}. In all the cases, the fully relativistic
analysis shows a clear improvement in describing the experimental
data compared with standard nonrelativistic treatments.

Although a treatment of FSI is necessary to describe 
experimental data, various studies have appeared in recent years
dealing with the relativistic plane-wave impulse approximation
(RPWIA). This simplified approach has proved to be very useful in
order to disentangle relativistic effects from distortion effects.
The unpolarized responses in $A(e,e'p)B$ reactions within RPWIA 
were already presented
in~\cite{cab98}. There important 
modifications with respect to the standard plane-wave impulse
approximation (PWIA) calculation were found 
due to the presence of negative-energy components in the 
relativistic bound nucleon wave function. The interference $TL$
response and asymmetry, $A_{TL}$, were shown to be very sensitive to
dynamical effects of relativity affecting the lower components.
These results persist in more realistic calculations including FSI. In
fact, data on $R^{TL}$ and $A_{TL}$ are a strong indication of the
crucial role played by dynamical relativistic effects~\cite{udi99,udi01}.

Let us recall that the differences between the fully relativistic
approach and the standard nonrelativistic one can be divided into 
kinematical and dynamical effects. The first are due to the
4-vector current operator, compared with the nonrelativistic one that usually
involves $\vec{p}/M_N$ expansions. The latter come from the
difference between the nucleon (bound and ejected) wave functions involved.
Within these dynamical relativistic effects one may distinguish
effects associated with the so-called Darwin term, that 
mainly affect the determination of spectroscopic factors at low 
missing momenta, and the ones due to the dynamical enhancement of 
the lower components of the relativistic wave functions,
which are expected to be more relevant at high 
missing momenta, although they might produce noticeable effects for
some particular observables even at low/medium $p_m$ values.
      
Our main aim in this paper is to study, within RPWIA,
the new response functions that enter in the analysis of
$A(\vec{e},e'\vec{p})B$ processes
due to the presence of spin-dependent 
degrees of freedom. All the results shown are for the
reaction $^{16}O(\vec{e},e'\vec{p})^{15}N$.
Following the arguments presented for the
unpolarized case in~\cite{cab98}, here we extend the analysis
to the polarized situation and for the polarized responses try to
identify clear signatures due solely to
the dynamical relativistic effects coming from the
negative-energy projections (NEP) of the relativistic bound nucleon wave 
function. Moreover, the role played by the NEP on the
transfer polarization components, $P'_l$, $P'_s$, is also
analyzed in detail. These issues are presented in Section~2. 
In Section~3 we also estimate the kinematical relativistic
effects introduced by using various possible non-relativistic
reductions of the nuclear current operator.
Finally in Section~4 we summarize our main conclusions.

\section{Dynamical relativistic effects}

As mentioned in the introduction, in this work we restrict our
attention to the RPWIA. Hence the ejected proton is described as a plane
wave, i.e., FSI are neglected.
Within this scheme, kinematical relativistic effects are
included via the use of the fully relativistic CC1 and/or CC2 current 
operators~\cite{for83}. Here we are mainly interested in the
dynamical relativistic effects coming from the presence of the 
negative-energy projections of the bound nucleon wave function
that gives rise to the dynamical enhancement of the lower components.

From the total of eighteen response functions that enter in the
analysis of $A(\vec{e},e'\vec{p})B$ reactions~\cite{pick8789},
only nine survive within RPWIA. Dynamical relativistic effects affecting
the four unpolarized responses ($R^L$, $R^T$, $R^{TL}$ and $R^{TT}$) 
were already studied in detail in~\cite{cab98}. Here we 
present results for the polarized response functions and transfer polarization 
components corresponding to two different selected kinematics:
\begin{enumerate}
\item ($q-\omega$) constant kinematics with $q=500$ MeV/c and 
$\omega=131.56$ MeV. The value of the transfer energy $\omega$ 
corresponds to the quasielastic peak value.
\item Parallel kinematics. The outgoing nucleon kinetic energy
is fixed at $120.2$ MeV, and the angle $\theta$ between the 
missing momentum $\vec{p}_m$ and the transfer momentum $\vec{q}$
is fixed at $0^o$ ($p_m>0$) and/or $180^o$ ($p_m<0$).
\end{enumerate}
In both cases coplanar kinematics ($\phi = 0^o$) has been selected and
we focus on the proton knockout from the $1p_{1/2}$ shell in $^{16}$O.
The bound state wave function has been computed within the Walecka 
relativistic model. It corresponds to a
Dirac-Hartree solution from a phenomenological
relativistic Lagrangian with scalar and vector meson potentials. 
The parameters of the set HS~\cite{hor7981} and the 
TIMORA code~\cite{hor91} have been used. Other
possible parameterizations of the bound state wave function have been
also checked. The results obtained show similar trends to the ones
presented in the figures that follow. For all the applications discussed below 
the Coulomb gauge has been chosen. 

Let us first discuss the results for the polarized responses. 
From the five polarized responses surviving within RPWIA, one of them,
$R^{TL'}_n$, only enters for out-of-plane kinematics. Therefore, in 
coplanar kinematics, the analysis is reduced to four polarized 
response functions: $R^{T'}_l$, $R^{TL'}_l$, $R^{T'}_s$ and
$R^{TL'}_s$. The
subindex refers to the three directions of the final-state polarization 
defined in the Introduction. The presence of the negative-energy 
components of the bound nucleon wave function breaks the 
factorization property that holds when only positive-energy components
are taken into account. Within RPWIA each response function can 
be decomposed in the form
\begin{equation}
\label{eq:resp}R^K ={\cal{R}}^k_PN_P(p)+{\cal{R}}^K_CN_C(p)+{\cal{R}}^K_NN_N(p).
\end{equation}
The first term is proportional to the square of the positive-energy 
projection, and is analogous to the standard result that appears in a
nonrelativistic calculation. The remaining two terms are proportional 
(quadratically and linearly) to the negative-energy projection 
(${\cal{R}}^K_NN_N(p)$ and ${\cal{R}}^K_CN_C(p)$).

\begin{figure}[htb]
{\par\centering \resizebox*{0.5\textwidth}{0.43\textheight}{\rotatebox{270}{\includegraphics{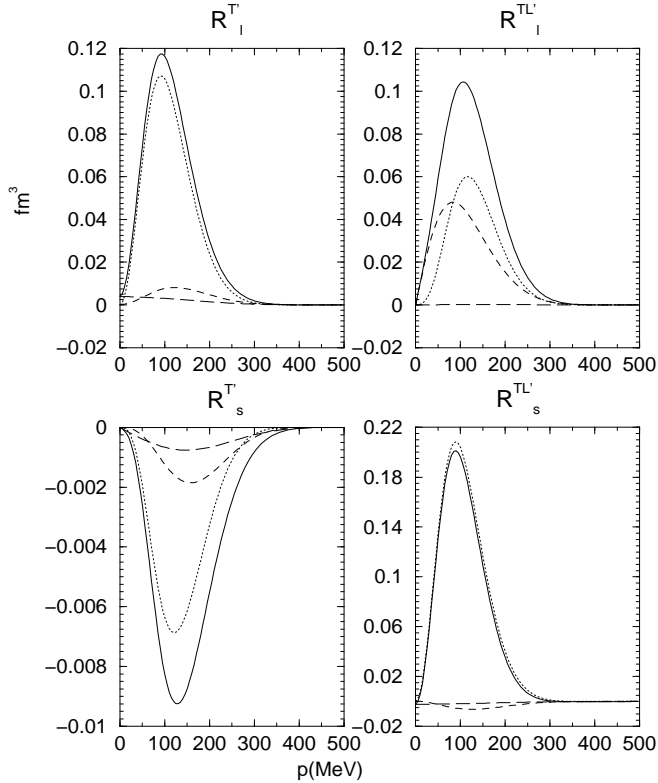}}} \par}
    \caption[]{\label{fig:percc10}
     Polarized hadronic responses for the $1p_{1/2}$ shell in $^{16}$O
     in ($q-\omega$) constant kinematics. Coulomb gauge and the CC1
     current operator have been used. Fully relativistic result (solid
     line) is compared to
its positive-energy component (dotted line), crossed one (short-dashed line) 
and negative-energy term (long-dashed line).}
\end{figure}

In Fig.~\ref{fig:percc10} we represent the results obtained for the
polarized response functions corresponding to 
($q-\omega$) constant kinematics and CC1 current operator. 
We show the fully relativistic result (solid) versus its three 
contributions as given in~(\ref{eq:resp}): the positive-energy (dotted),
the crossed (short-dashed) and the negative-energy 
(long-dashed) contributions. 
Thus the dynamical relativistic effects are easily
appreciated by just comparing the
negative-energy and crossed terms with the total response. It is clearly seen
that in two responses, $R^{T'}_l$ and $R^{TL'}_s$, the contribution 
of NEP is almost negligible, that is, dynamical relativistic effects 
from the bound nucleon wave function do not affect these responses. 
On the contrary, the two remaining polarized responses are very 
sensitive to these effects. In both cases, although
the negative-energy term does not contribute significantly to the total
result, the crossed term plays an important role, especially
for the $R^{TL'}_l$ response where its contribution is similar to the
one coming from the positive-energy projection.
This result resembles what appeared for the unpolarized interference
$TL$ response. Hence there exists
a strong discrepancy between RPWIA results and those corresponding to
the standard PWIA (we must recall that although the positive-energy term 
in~(\ref{eq:resp}) is not identical to the PWIA result, for which 
we must take the nonrelativistic momentum distribution $N_{nr}(p)$, 
the difference is very small provided that $N_{nr}(p)\sim N_P(p)$).

\begin{figure}[htb]
{\par\centering \resizebox*{0.5\textwidth}{0.43\textheight}{\rotatebox{270}{\includegraphics{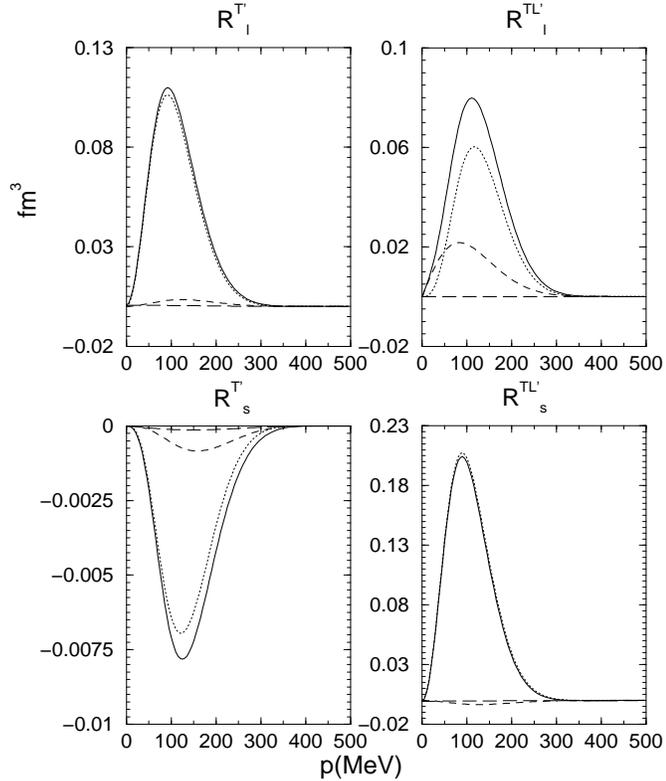}}} \par}
    \caption[]{\label{fig:percc20}
     Polarized hadronic responses for the $1p_{1/2}$ shell in $^{16}$O
     in ($q-\omega$) constant kinematics. Coulomb gauge and the CC2
     current operator have been considered. The labelling is the same 
     as in Fig.~\ref{fig:percc10}.}
\end{figure}

In Fig.~\ref{fig:percc20} we show the polarized responses obtained
using the CC2 form of the nucleon current operator. As observed, the
general trend is similar to the one discussed for the CC1 case, 
except for the magnitude of the relativistic effects. Although
the role of the negative-energy and crossed terms is significantly reduced for
the CC2 current (Fig.~\ref{fig:percc20}), their effects
are still quite sizeable on $R^{T'}_s$ and $R^{TL'}_l$. 
This behaviour is similar to the one already stated for the 
unpolarized responses in~\cite{cab98}. The use of the CC1 operator 
maximizes the role of the negative-energy projections. This can be 
traced back to the fact that the CC2 form of the current is
obtained (for free nucleons) by simply imposing general constrains 
over the more general form of the current. On the contrary,
the CC1 operator is obtained from the CC2 one by applying the 
Gordon decomposition, only valid for free $u$ Dirac spinors. Note that
in RPWIA one should also take into account couplings to $v$ Dirac spinors
for which the Gordon decomposition is not valid. This explains why the
difference between CC1 and CC2 results is much more important for the
negative-energy and crossed terms than for the strictly 
positive-energy contribution (standard PWIA calculations).

\begin{figure}[htb]
{\par\centering \resizebox*{0.55\textwidth}{0.35\textheight}{\rotatebox{270}{\includegraphics{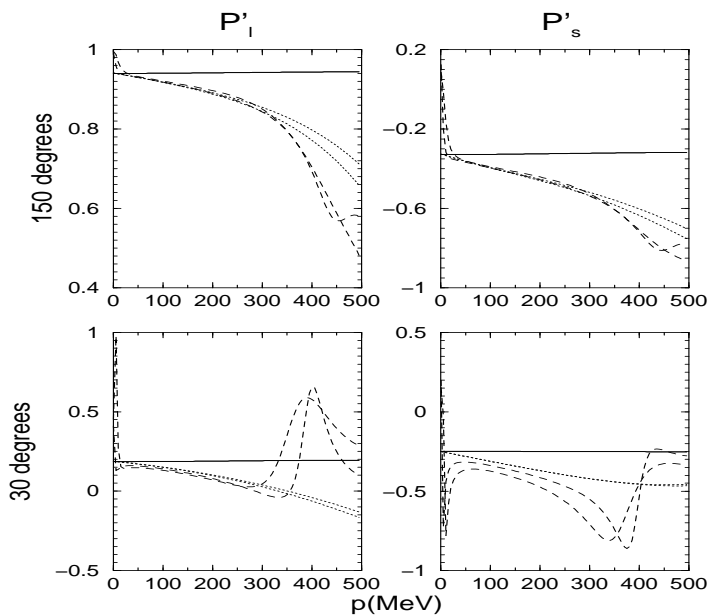}}} \par}
    \caption[]{\label{fig:perpol}
     Longitudinal and sideways transferred polarizations for the 
$1p_{1/2}$ shell in $^{16}$O in ($q-\omega$) constant kinematics. 
Coulomb gauge and CC1 (thin lines) and CC2 (thick lines) currents are 
considered.
Top panels correspond to $\theta_e=150^o$ and bottom panels to 
$\theta_e=30^o$. The fully relativistic results are represented by 
dashed lines while their projected results are given by the dotted
lines. Also shown as a guide is the result for free
electron-nucleon scattering (solid line).}
\end{figure}

In what follows we analyze the behaviour of the transfer polarization
asymmetries $P'_l$ and $P'_s$. Note that the ratio
$P'_s/P'_l$ is related to the nucleon electric/magnetic form factors.
In Fig.~\ref{fig:perpol} we show the results corresponding to
forward ($\theta_e=30^o$) and backward ($\theta_e=150^o$) electron
scattering angles. In the former case the electron beam energy is
given by $\varepsilon_e=1$ GeV and in the latter $\varepsilon_e=324$ MeV.
We compare the fully relativistic results (dashed lines) corresponding
to the CC1 and CC2 current operators with their positive-energy projection 
contributions (dotted lines). The difference between the relativistic 
and projected results observed for very small missing momentum values 
is directly connected to the quantum number $\overline{\ell}=0$ of the
lower component in the $1p_{1/2}$ state
(see ref.~\cite{cab98} for details). 
Apart from this behaviour for very small missing momenta, it is
important to note that fully relativistic and positive-energy projected
results do not differ appreciably (especially for backward angles)
for $p_m$-values up to $\sim 300$ MeV/c. For $p_m > 300$ MeV/c
both (relativistic vs projected) results start to deviate from each
other. This general behaviour is what one should expect because of
the clear dominance of the positive-energy projection component of the momentum
distribution in the region $p_m\leq 300$ MeV/c~\cite{cab98}.
On the contrary, in the region of high missing momentum, $p_m>300$
MeV/c, the crossed and negative-energy components, $N_C(p), N_N(p)$, are
similar to or even larger than that of $N_P(p)$, hence the effects of the
dynamical enhancement of the lower components in the bound
relativistic wave function are clearly visible in the transfer 
polarization asymmetries.

Finally, it is also clear from the results shown in
Fig.~\ref{fig:perpol} that the dynamical effects are maximized in the
forward electron scattering situation. Here the differences between
fully relativistic and projected results are important even for
low/medium $p_m$ values, in particular for the sideways transfer
polarization, $P'_s$. 
In Table~\ref{tab:vk} we present the various electron kinematical
factors that enter in the polarized cross section. We display the
values corresponding to forward ($\theta_e=30^o$) and 
backward ($\theta_e=150^o$) kinematical situations. As noted,
the purely transverse responses dominate at backward angles; hence
the most relevant contributions to the polarization asymmetries
in this case come from the transverse polarized responses $R^{T'}_l$ 
and/or $R^{T'}_s$ in the numerator, and 
from the unpolarized $R^T$ response in the denominator.
From these three responses, only the small $R^{T'}_s$
is particularly sensitive
to the effect of the negative-energy components.
On the contrary, at a forward angle ($\theta_e=30^o$)
all the kinematical factors (table I) are of similar order, and hence
the contribution of the responses 
that are more sensitive to dynamical relativistic effects 
is clearly maximized. Therefore, the important role played by the 
negative-energy components of the bound relativistic wave function is much
better appreciated for the transfer polarization asymmetries measured
at forward angles. 

\begin{table}
{\centering \begin{tabular}{ccccccc}
\hline 
\( \theta (^{o}) \)&
\( v_{L} \)&
\( v_{T} \)&
\( v_{TL} \)&
\( v_{TT} \)&
\( v_{T'} \)&
\( v_{TL'} \)\\
\\
\hline 
\\
30&
0.866&
0.537&
-0.659&
-0.465&
0.268&
-0.176\\
\\
\hline 
\\
150&
0.866&
14.39&
-2.537&
-0.465&
14.386&
-2.456\\
\\
\hline 
\end{tabular}\par}

\caption{\label{tab:vk}Electron kinematical coefficients \protect\( v_{k}\protect \) for 
perpendicular kinematics and two values of the electron scattering angle. }
\end{table}

\begin{figure}[htb]
{\par\centering \resizebox*{0.6\textwidth}{0.25\textheight}{\rotatebox{270}{\includegraphics{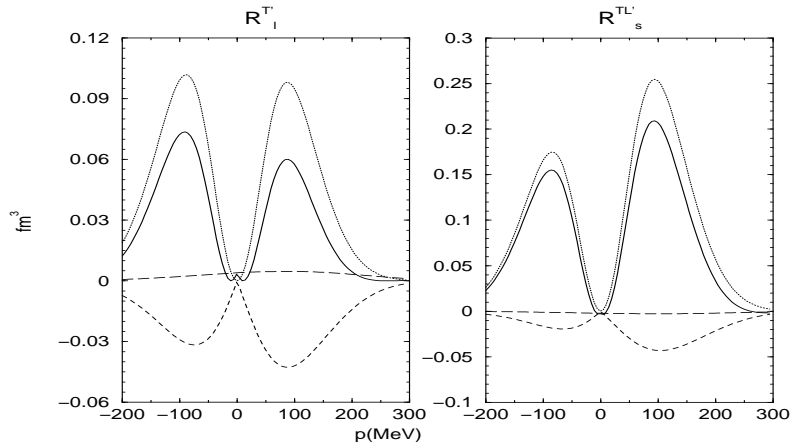}}} \par}
    \caption[]{\label{fig:parcc10}
     Polarized hadronic responses for the $1p_{1/2}$ shell in $^{16}$O
     in parallel kinematics. Coulomb gauge and the CC1 current
     operator have been considered.
     Labelling as in Fig.~\ref{fig:percc10}.}
\end{figure}

In parallel kinematics only two polarized responses survive: 
$R^{T'}_l$ and $R^{TL'}_s$. In Fig.~\ref{fig:parcc10} we present
both responses in parallel kinematics comparing the fully relativistic
result (solid line) with the positive-energy (dotted), crossed 
(short-dashed) and negative-energy (long-dashed) contributions.
Coulomb gauge
and CC1 current operator have been chosen. Note that in
($q-\omega$) constant kinematics these responses were the
less sensitive
to dynamical relativistic effects coming from the 
NEP of the bound wave function. However, in parallel kinematics
(Fig.~\ref{fig:parcc10}) both responses are clearly more sensitive to
dynamical relativistic effects than in the previous case.
Moreover, contrary to the ($q-\omega$) constant kinematics, where the
role of the negative-energy components was to increase the fully
relativistic result compared with the positive-energy projection, in
parallel kinematics the crossed term gives a negative contribution,
hence diminishing the fully relativistic response.
From these results it seems to be clear that parallel kinematics
enhances the sensitivity to
dynamical relativistic effects of the two surviving polarized responses
compared with these same responses evaluated in perpendicular kinematics.
Finally, we do not show here results for the transfer polarizations
in parallel kinematics: the outgoing nucleon kinetic energy
selected only allows us to reach low/medium missing momentum values
for which dynamical effects are much less relevant.

\section{Kinematical relativistic effects}

For a long time the standard procedure to treat $(e,e'p)$ reactions has
been based on a nonrelativistic description of the hadronic current.
Bound nucleon wave functions have been described as solutions of the 
Schroedinger equation and a nonrelativistic treatment of FSI has been
considered in describing the ejected nucleon wave function. The main 
reason for this analysis is directly connected with
the fact that most nuclear models have been derived within a 
nonrelativistic framework. Hence, in order to be consistent with such 
description of the nucleus, one is also forced to perform a 
nonrelativistic reduction of the relativistic electromagnetic current. 

In the previous section we have analyzed the dynamical relativistic 
effects coming from the bound state. In doing that we have compared the fully 
relativistic results obtained within RPWIA with the positive-energy
projected calculations. In this section we focus on the kinematical 
relativistic effects. In order to disentangle clearly both types of
relativistic effects and simplify the discussion we
only retain the positive-energy component of the responses. 
Thus by comparing these projected results with those evaluated by
performing a nonrelativistic reduction one would be able to estimate 
the magnitude of the relativistic kinematical effects.  
 
The standard nonrelativistic procedure has been based on expansions
in all dimensionless momenta, i.e., nonrelativistic expansions of 
the current were made in powers of the transferred momentum,
$\frac {q}{M_N}$, transferred energy, $\frac {\omega}{M_N}$, 
and momenta of the initial-state struck nucleons, $\frac {p_m}{M_N}$.
These approximations are not justified in present experiments,
since values of $q$ can be even higher than the nucleon mass.
Here, for comparison with experimental data, we treat the problem
exactly for the transferred energy and momentum, considering
only expansions in powers of $\frac{p_m}{M_n}$. Analyses of the 
nonrelativistic reductions along this line can be found in the 
literature~\cite{ama96,jes98}. In this work, instead of making use 
of existing nonrelativistic expressions for the
single-particle current matrix elements, we directly expand the 
single-nucleon responses in powers of $\frac{p_m}{M_N}$ up to first
order. We compare the results so obtained
to the fully relativistic PWIA calculation. Our main aim
is to establish how precise the expansion in powers of 
$\frac{p_m}{M_N}$ is, and under which conditions and/or for which observables 
it does or does not work. In what follows we only analyze the transfer 
polarization asymmetries which, within PWIA, do not depend on 
the nuclear model, since factorization is fulfilled.

\begin{figure}[htb]
{\par\centering \resizebox*{0.55\textwidth}{0.35\textheight}{\rotatebox{270}{\includegraphics{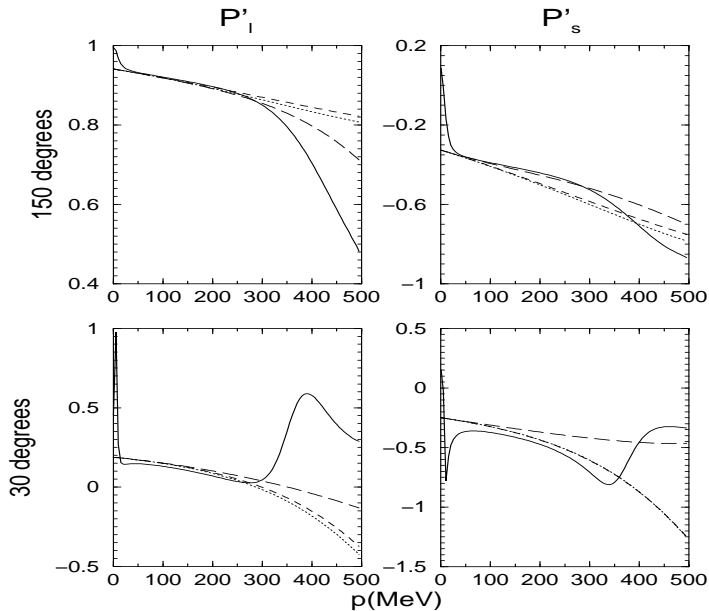}}} \par}
    \caption[]{\label{fig:kinperpol}
     Longitudinal and sideways transfered polarizations for the 
$1p_{1/2}$ shell in $^{16}$O in ($q-\omega$) constant kinematics, 
using the Coulomb gauge and the CC1 current. Top panels correspond to 
$\theta_e=150^o$ and bottom panels to $\theta_e=30^o$. 
The fully relativistic results are represented with solid lines, 
the projected results are given by long-dashed lines 
and the other two curves (dotted and shot-dashed) correspond to two 
different nonrelativistic reductions.}
\end{figure}

In Fig.~\ref{fig:kinperpol} we represent two different nonrelativistic
results (dotted and short-dashed lines) corresponding to 
the ($q-\omega$) constant kinematics (see ref.~\cite{cris1} for
details). Fully relativistic (solid line) and positive-energy projected 
(long-dashed line) calculations are also shown
for comparison. The kinematical effects are clearly visible by
comparing the nonrelativistic curves to the projected ones. We observe
that kinematical effects are almost negligible in the low $p_m$
region ($p\leq 300$ MeV), starting to show up for higher missing momenta.
Note that this was also the case for the
dynamical effects. Moreover, kinematical relativistic effects are
also maximized at forward electron scattering angle (bottom panels). 
Finally, it is important to remark that the use of nonrelativistic 
reductions may even produce 
unphysical results: the sideways transferred polarization, $P'_s$,
for $\theta_e=30^o$ gets bigger than 1 (in absolute value) for 
high $p_m$ which means negative cross section. Therefore, one should
be very careful in doing non-relativistic expansions to use in the
analysis of present-day experiments.

\section{Summary}

Dynamical relativistic effects associated with the bound nucleon wave
function have been analyzed within RPWIA for polarized hadronic
responses and transferred polarization observables. Two different
kinematical situations have been considered:
i) ($q-\omega$) constant kinematics and ii) 
parallel kinematics. We have
found that the four polarized responses that enter in the analysis
of coincidence electron scattering reactions may be quite sensitive to
the presence of negative-energy projections in the relativistic bound 
state. In particular,
\begin{itemize}
\item $R^{T'}_s$ and $R^{TL'}_l$ show important deviations from the 
positive-energy projected result in ($q-\omega$) constant kinematics, 
whereas the role of NEP on the two remaining polarized responses is almost
negiglible.
\item $R^{T'}_l$ and $R^{TL'}_s$ are, however, particularly sensitive to
  NEP in the case of parallel kinematics.
\end{itemize}
 
Concerning the transfer polarizations, it should be pointed out that
they can be strongly affected by the negative-energy contributions, 
mainly at high $p_m$-values. In particular, at a forward 
electron scattering angle
($\theta_e=30^o$) the dynamical enhancement of the lower components in
the bound nucleon wave function may modify completely the structure of the
polarization asymmetries. 
Kinematical effects can also be very important under the same conditions, 
and we must keep in mind their limits of validity at high missing
momenta. This region is particularly interesting if one wishes
to investigate short-range correlations. 

Finally, although being aware of the important modifications that FSI 
may introduce in the analysis, we are rather confident that
the high sensitivity of polarization-related observables 
to negative-energy projections already shown within RPWIA, will be
probably also maintained within more elaborated relativistic
distorted-wave impulse approximation (RDWIA) calculations. 
Work along this line is presently in progress.

\section*{Acknowledgments}
M.C.M. acknowledges a grant from the Fundaci\'on C\'amara of the Universidad de Sevilla.

\end{document}